\documentclass{pasj01}

\usepackage{color}

\usepackage{ulem}

\Received{2015 June 3}
\Accepted{2016 January 28}
\Published{2016}
\jyear{2016}

\begin{document}

\title{Theory of the Jitter radiation in a magnetized plasma accompanying temperature gradient}

\author{M.Hattori and K.Fujiki}
\affil{Astronomical Institute, Graduate school of science, Tohoku University,
    Sendai 980-8578, Japan}

\email{hattori@astr.tohoku.ac.jp}

\maketitle

\begin{abstract}

The linear stability of a magnetized plasma accompanying temperature gradient was reexamined by using plasma kinetic theory. 
The anisotropic velocity distribution function was decomposed into two components. 
One is proportional to the temperature gradient parallel to and the other is proportional to the temperature gradient perpendicular to the back ground magnetic field. 
Since the amplitude of the anisotropic velocity distribution function is proportional to
the heat conductivity and the heat conductivities perpendicular to the magnetic field is strongly reduced,  
the first component of the anisotropic velocity distribution function is predominant. 
The anisotropic velocity distribution function induced by the temperature gradient 
along the back ground magnetic field drives plasma kinetic instability
and the circular polarized magnetic plasma waves are excited. 
The instability is almost identical to Weibel instability in weakly magnetized plasma. 
However, depending on whether wave vectors of modes 
are parallel to or antiparallel to the back ground magnetic field, the growth rate is suppressed or enhanced due to back ground magnetic field.  
In the strongly magnetized plasma, one mode is stabilized and only one of the modes remains unstable. 

The Jitter radiation spectrum formulae emitted by relativistic electrons 
when they travel through the magnetized plasma with the plasma waves driven by 
the instability, are deduced at the first time. 
The synchrotron emission and the Jitter radiation are simultaneously emitted 
from the same relativistic electron. 
The Jitter radiation is expected to be circularly polarized but 
with a very small polarization degree 
since almost the same amount of  left and right handed circular polarized magnetic waves are excited by
the instability.

\end{abstract}


\section{Introduction}

Astronomical roles of the Weibel types instabilities have examined since its discovery (\cite{Wei1959}).
The Weibel instability has been 
proposed as a generation mechanism of the magnetic fields in astronomical shock waves (\cite{MedLoeb1999}, \cite{Hunt2015}),
of the cosmological magnetic from zero seed field (\cite{SchShuk2003}),
and of origin of the interstellar turbulent field (\cite{Tautz2013}). 
It has been also known that the temperature gradient of the thermal electrons 
drives the Weibel type plasma kinetic instability   (\cite{Wei1959}, \cite{RL1978}, \cite{OH2003}). 
By following \cite{OH2003}, we refer this instability as the RL instability since this instability was 
first found by \cite{RL1978}. 
The RL instability has been studied as a mechanism of the reduction of the heat conductivities in 
the intracluster medium (\cite{Pist1996}, \cite{HatUme2000})
and as a mechanism to maintain the sharp interface of the cold fronts found in the intracluster medium 
(\cite{OH2003}). 
However, it has been yet unclear whether the RL instability plays significant role in some astronomical situation. 
Therefore, the studies on the observational tests to judge whether the RL instability really plays a central role 
in the generation of the magnetic field or not are desired. 
Spectrum distortion of the comic microwave background radiation caused by the 
inverse Compton scattering by thermal electrons which belongs 
to the anisotropic electron velocity distribution function generated by the 
temperature gradient proposed by \cite{Hattori2005} could be one of such tests. 

The Jitter radiation emitted from the relativistic electrons when they travel through the 
magnetic fields generated by the Weibel instability has been studied as an alternative 
interpretation of the gamma ray burst after glow (\cite{Med2000}). 
In the case of the RL instability, the Jitter radiation is also expected to be observed.  
The typically wavelength of the excited waves are order of the plasma penetration length, that is 
$c/\omega_{pe}$, 
where $c$ is the speed of light and $\omega_{pe}$ is the plasma frequency defined by
$\omega_{pe}=\sqrt{4\pi e^2 n_e/m_e}$ where $-e$ is a charge of an electron, $m_e$ is an electron mass
and $n_e$ is an electron number density. 
When the saturation level of the amplitude of the magnetic fields in the excited waves are defined by the thermal electron 
trapping by the magnetic field of the excited waves, 
the amplitude of the magnetic field in the excited waves is described by 
$\omega_{ce1}\sim (v_{th}/c)\omega_{pe}$ where $v_{th}$ is the 
electron thermal velocity defined by $\sqrt{2k_B T_e/m_e}$, $k_B$ is the Boltzmann constant, $T_e$ is the electron 
thermal temperature and 
$\omega_{ce1}=eB_1/m_e c$ is the electron cyclotron frequency 
defined by the magnetic field strength of the excited waves, that is denoted by $B_1$. 
Orbits of relativistic electrons are perturbed by the magnetic field of the excited waves. 
The deflection angle, $\theta_d$,  when the relativistic electron with the Lorentz factor of $\gamma$ 
travels through the excited waves, is about the ratio between the wave length of the excited plasma waves 
and the Larmor radius for the relativistic electron, that is  $\theta_d\sim (1/\gamma)(\omega_{ce1}/\omega_{pe})\sim
(1/\gamma) (v_{th}/c)$. It shows that, as far as non relativistic thermal plasma is considered, 
the deflection angle is much less than the angle spanned by the relativistic beaming cone, that is $1/\gamma$.
Once the line of sight is included in the relativistic beaming cone, the line of sight is continuously included in it
during the relativistic electron travel through the excited waves. 
The emission caused by this situation is the Jitter radiation (\cite{Med2000}). 
The observed frequency of the Jitter radiation in this situation is 
defined by the Doppler shifted plasma frequency, that is $\gamma^2 \omega_{pe}$. 

The extension of the above studies to the magnetized plasma is required 
when we  apply the results to astrophysical situation since almost all the 
astrophysical plasma are magnetized. 
The plasma kinetic instability driven by electron temperature gradient in a magnetized plasma 
was examined by \cite{LE1992}.
They found that the unstable modes driven by the temperature gradient exist even in the magnetized plasma. 
They also showed that there are significant differences of the behaviors of the instability between 
weakly magnetized plasma and strongly magnetized plasma. 
\cite{OH2003} has shown that the physical essence of the RL instability in unmagnetized plasma
 is the same as the Weibel instability. 
Studies on the Weibel instability in a magnetized plasma were done by \cite{Lazar2009}. 
The comparison of the results of the RL instability in a magnetized plasma 
with the results obtained by \cite{Lazar2009} may help to deepen our understandings 
of the physics of the RL  instability in a magnetized plasma. 
Because the physical similarities between the RL and the Weibel instabilities were not clearly recognized 
before \cite{OH2003}, 
the comparison was not done by \cite{LE1992}. 
In this paper, we summarize the characteristics of the RL instability in a magnetized plasma
by comparing the Weibel instability. 
More over, \cite{LE1992} did not show the nature of the eigen modes. 
However, to analyze radiative processes when relativistic electrons travel through the 
excited waves, the nature of the eigen modes are important. 
Therefore, we show the eigen modes of the unstable modes.  
\cite{Med2000} studied the Jitter radiation when the relativistic electrons travel through the 
Weibel turbulence. 
However, the radiative processes in the situation  when the ordered  magnetic field and the wave magnetic field 
generated by the RL instability in a magnetized plasma  coexist, have not yet been done. 
Therefore, we study the radiative processes in the situation  when the ordered magnetic field and the wave magnetic field 
generated by the RL instability in a magnetized plasma  coexist, 
and deduce the spectrum formulae of the Jitter radiation in this situation. 

The Jitter radiation has been studied for years  (e.g. \cite{LL1975}, \cite{TF1987}, \cite{RK2010}, \cite{TT2011}, \cite{KAK2013}).
The emission of electromagnetetic waves due to the interaction of relativistic electrons with regular and 
turbulent magnetic fields were first studied by \cite{TF1987}.  
The emission arised by the regular field is characterized by the synchrotron emission
and the emission raised by the small scale turbulent field is characterized by the Jitter radiation. 
They showed that 
the frequency of the Jitter radiation is characterized by $\gamma^2 (c/\ell_{min})$ 
and much higher than the characteristic frequency of the synchrotron emission 
emitted by the same relativistic electron when the minimum length of 
the coherent scale of the turbulence, $\ell_{min}$,  is much shorter than the non relativistic Larmor radius $R_L=c/\omega_{ce0}$ where $\omega_{ce0}$ is the electron cyclotron frequency defined by the regular magnetic field.
\cite{KAK2013} presented the general formulae of the radiation within the frame work of perturbation theory 
and restressed the distinct spectral features of the Jitter radiation in which the emitted radiation shifted toward higher energy compared to synchrotron emission. 
\cite{TF1987} showed that both synchrotron and Jitter radiations appeared in the frequency range less than 
$\gamma^2 \omega_{ce0}$. 
However, their turbulent magnetic field models did not have certain physical grounds. 
In this paper, we have linked the RL instability in a magnetized plasma to small scale turbulent magnetic fields
and performed self consistent studies of the radiation emitted  by relativistic electrons at the first time. 
Numerical studies of the Jitter radiation based on the particle-in-cell technique (\cite{RK2010}, \cite{TT2011})
has been progressed. 
Although this method has a great potential to deal with complex situations, 
their studied cases were limited in the situations where $\ell_{min} > R_L$ up to now
which are outside of a range with our current interests.

The structure of the paper is as follows. 
In section 2, we revisit plasma kinetic instability driven by electron temperature gradient in a magnetized plasma. 
In section 3, we present theory of the Jitter radiation in a magnetized plasma when the relativistic electrons travel 
through the magnetic fields associated with the plasma waves excited by the instability. 
Section 4 is dedicated to the discussions.

\section{Plasma kinetic instability driven by electron temperature gradient in a magnetized plasma}

In this section, plasma kinetic instability driven by electron temperature gradient in a magnetized plasma are summarized. 
The wavelengths of the modes which we are interested in, are much smaller than the mean free path of thermal electron.
Therefore, the plasma kinetic theory is used to analyze the linear stability (\cite{Ichi1973}).
The driving force of the instability is anisotropy of thermal electron velocity distribution function induced by 
temperature gradient (\cite{RL1978}).
The amplitude of the deviation of the  electron velocity distribution function from the Maxwell Boltzmann distribution 
is characterized by relative variation of the electron temperature, $\delta_T=\delta T_e/T_e$, and
ratio between electron mean free path
 and a scale of electron temperature variation,  
$\epsilon=\lambda_e/\delta x$, where we assume that the electron temperature 
varies $\delta T_e$ across the scale $\delta x$ along the temperature gradient.

The anisotropic electron velocity distribution function which is the driving force of the plasma kinetic instability, 
is obtained by expanding electron velocity distribution function perturbatively in $\epsilon\delta_T$ following the procedures of the Chapman-Enskog expansion (\cite{ChCow}).
In an unmagnetized plasma, the anisotropic part of the 
electron distribution function up to the first order in $\epsilon\delta_T$ is described by 
\begin{eqnarray}
f^{(1)}&=&\epsilon\delta_T {v_{\parallel\nabla T_e}\over v_{th}}\left({5\over 2}-{v^2\over v_{th}^2}\right)f_m,\label{eq:ChapEns}
\end{eqnarray}
where $f_m$ is the Maxwell-Boltzmann electron velocity distribution function,
$v_{\parallel\nabla T_e}$ is the velocity component along the temperature gradient (\cite{RL1978}, \cite{OH2003}). 
The anisotropic velocity distribution is directly related to the heat current density as 
\begin{eqnarray}
-\kappa_{\rm{Sp}} \vec{\nabla}  T_e&=&\int d^3 \vec{v} {m_e v^2\over 2} \vec{v} f^{(1)}, \label{eq:Hcond}
\end{eqnarray}
where $\kappa_{\rm Sp}$ is the Spitzer heat conductivity  (\cite{Hattori2005})
and we neglect a factor of order one deviation.  
The Spitzer heat conductivity is given by $\kappa_{\rm Sp}\sim \lambda_e n_e k_B v_{th}$ (\cite{Sp1956}, \cite{Sarazin1988}).
Therefore, the anisotropic electron velocity distribution function is rewritten by using $\kappa_{\rm Sp}$ as
\begin{eqnarray}
f^{(1)}&\sim& \kappa_{\rm Sp} {1\over k_B n_e} {\vec{v}\cdot\vec{\nabla} T_e \over v_{th}^2 T_e} \left({5\over 2}-{v^2\over v_{th}^2}\right) f_m.
\label{eq:DfSp}
\end{eqnarray}
Since the heat conductivities in the directions of  perpendicular to the magnetic field, $\kappa_{\perp}$,  
are dramatically reduced (\cite{Sp1956}) from the Spitzer value, 
we propose that the anisotropic electron velocity distribution function in a magnetized plasma is 
decomposed into following two parts as 
\begin{eqnarray}
f^{(1)}&\sim& \kappa_{\rm Sp} {1\over k_B n_e} {v_{\parallel} \nabla_{\parallel} T_e \over v_{th}^2 T_e} \left({5\over 2}-{v^2\over v_{th}^2}\right) f_m\nonumber\\
&&+\kappa_{\perp} {1\over k_B n_e} {v_{\perp} \nabla_{\perp} T_e \over v_{th}^2 T_e} \left({5\over 2}-{v^2\over v_{th}^2}\right) f_m,
\label{eq:Dfparaperp}
\end{eqnarray}
where $v_{\parallel}$ and $v_{\perp}$ are electron velocity components parallel to and perpendicular to magnetic field, 
and the temperature gradients parallel and perpendicular to the magnetic field are denoted by $\nabla_{\parallel} T_e$ 
and $\nabla_{\perp} T_e$, respectively. 
Since $\kappa_{\perp}\ll \kappa_{\rm Sp}$, 
we adopt the following approximated form as for the anisotropic electron velocity distribution function in a magnetized plasma as
\begin{eqnarray}
 f^{(1)}&=& \delta f_1 {v_{\parallel}\over v_{th}}\cos \theta_B \left({5\over 2}-{v^2\over v_{th}^2}\right)f_m,
\label{eq:Df}
\end{eqnarray}
where $\delta f_1\sim \epsilon\delta_T$ and 
$\theta_B$ is an inclination angle of the temperature gradient relative to the back ground magnetic field. 
In the following discussion, we drop a factor of $\cos \theta_B$ appeared in Eq.(\ref{eq:Df}) 
since $\cos \theta_B$ stays order of one except $\theta_B$ has a value very close to 
$\pi/2$. 

The linear stability of the RL instability in the magnetized plasma was first studied by \cite{LE1992}.
In the rest of this section, we summarize the characteristics of  the RL instability in a magnetized plasma 
by comparing with the characteristics of the Weibel instability in a magnetized plasma obtained by \cite{Lazar2009} aiming to 
help physical understandings of the RL instability in a magnetized plasma.
As shown by \cite{RL1978}, the modes which have wave vectors nearly perpendicular to 
the temperature gradient are stable even in the absence of the back ground  magnetic field. 
Further, the amplitude of the anisotropy of the electron velocity distribution function due to the 
temperature gradient which is nearly perpendicular to the back ground magnetic field is
negligibly small as discussed in above. 
Therefore, we may safely conclude that the modes with wave vectors which are parallel or antiparallel to 
the back ground magnetic field  are the most relevant to the RL instability in a magnetized plasma (\cite{LE1992}). 
In the following discussion, we take $k_{\perp}\rightarrow 0$ limit where $k_{\perp}$ is the 
component of the wave vectors perpendicular to the magnetic field. 
In this limit, only two fundamental modes with $n=+1, -1$ are predominant against the higher order  harmonics
({\cite{Ichi1973}). 
The dispersion relation is described by the function of $\xi_{\pm}\equiv (\omega\pm\omega_{ce})/k v_{th}$
where the double sign same order and each sign corresponds to $n= \mp 1$, respectively. 
Characteristics of the unstable modes show the different behaviors for small and large $|\xi|$ limits. 

First the results for $|\xi|\ll 1$ and $\delta f_1\ll 1$ are summarized.  
Imaginary parts of the angular frequency are given by 
\begin{eqnarray}
\omega_i^{(n)}&\sim & {1\over \sqrt{\pi}}\left( {3\over 8} \delta f_1^2 k v_{th}-{c^2\over \omega_{pe}^2} k^3 v_{th}\mp {\delta f_1\over 2} n \omega_{ce}\right), \label{eq:omiWeiblim}
\end{eqnarray}
where $k= |\vec{k}|$, the last term takes upper sign, that is $(-)$,  when $\vec{k}$ is parallel to the back ground magnetic field, $\vec{B}_0$,  and 
lower sign, that is $(+)$, when $\vec{k}$ is antiparallel to the back ground magnetic field. 
It takes the maximum value at $k=k_m= {1\over 2\sqrt{2}}\delta f_1 \omega_{pe}/c$. 
Real parts of the angular frequency are given by 
\begin{eqnarray}
\omega_r^{(n)}&\sim & \pm {1\over 4}\delta f_1 k v_{th} \mp {2\over \sqrt{\pi}} \delta f_1 \omega_i,\nonumber\\
&\sim&  \pm {1\over 4}\delta f_1 k v_{th} +{1\over \pi} \delta f_1^2 n \omega_{ce},\label{eq:omrWeiblim}
\end{eqnarray}
where the first term of the last equation takes upper or lower sign when $\vec{k}$ is parallel to or antiparallel to the back ground 
magnetic field, respectively. 
In Eq.(\ref{eq:omrWeiblim}), term higher than third order in $\delta f_1$ are neglected and $k\sim k_m$
is assumed.  As far as the absolute values of the wave vectors stay around  $k\sim \delta f_1 \omega_{pe}/c$, 
\begin{eqnarray*}
|\xi_r|&\sim& \delta f_1 +\beta^{-1/2} \delta f_1,\\
|\xi_i|&\sim& \beta^{-1/2} +\delta f_1^2,
\end{eqnarray*}
where   the plasma $\beta$ is defined by the ratio of thermal pressure to the back ground magnetic field pressure 
as $\beta= 2 n_ek_BT_e/(B_0^2/8\pi)$ where the mean molecular weight is 
set to be 0.5 which corresponds to the fully ionized pure hydrogen plasma.  
The plasma $\beta$ is expressed by a combination of the electron thermal velocity, $v_{th}$,  the 
  electron cyclotron frequency for the back ground magnetic field, $\omega_{ce0}$,  
  and the plasma frequency as $\omega_{ce0} c/(\omega_{pe} v_{th})=(2/\beta)^{1/2}$.
Therefore, small $|\xi|$ limit requires \sout{that} high $\beta$, that is $\beta\gg 1$, and small amplitude of the anisotropy, that is $\delta f_1\ll 1$.  
The first term of Eq.(\ref{eq:omrWeiblim}) exactly coincides with the real part of the RL instability for unmagnetized plasma (\cite{OH2003}).  The second term of  Eq.(\ref{eq:omrWeiblim}) is almost equivalent to the equation (18) in \cite{Lazar2009}.
This result is natural consequence since the RL instability for unmagnetized plasma is essentially equivalent to the 
Weibel instability with relative directional temperature difference, parameter $A$ in \cite{Lazar2009}, of $\delta f_1^2$ (\cite{OH2003}).  
The first and second terms of the imaginary parts exactly coincide with the imaginary parts of the RL instability for unmagnetized plasma (\cite{OH2003}). 
The last term in Eq.(\ref{eq:omiWeiblim}) is uniquely appeared in the case of the RL instability in a magnetized plasma
and first found by \cite{LE1992} although the physical origin of the appearance of this term has not been clarified yet. 
When $\vec{k}$  is parallel to $\vec{B}_0$ and $n=+1$, the RL instability is stabilized by the existence of the magnetic fields and
$\beta > \delta f_1^{-4}$ is required for the modes to be unstable. 
On the other hand, when $\vec{k}$ is antiparallel to $\vec{B}_0$ and $n=+1$, the existence of the magnetic fields make the modes
more unstable. 
For the modes with $n=-1$, situations become vice versa. 
The most unstable modes appear at $k=k_m$  when $\vec{k}$ is antiparallel to $\vec{B}_0$ for $n=+1$ and 
$\vec{k}$ is parallel to $\vec{B}_0$ for $n=-1$.  The growth rates of these maximum growth rate modes are given by
\begin{eqnarray}
\omega_{i,m}&=& {\delta f_1 \over \sqrt{\pi}}\left({\delta f_1^2\over 8\sqrt{2}} {v_{th}\over c} \omega_{pe}+{1\over 2} \omega_{ce}\right),
\label{eq:grmWeiblim}
\end{eqnarray}
and real parts of the angular frequency of these modes are given by 
\begin{eqnarray}
\omega_{r,m}^{(\pm)}&=& \delta f_1^2\left( \pm{1\over 8\sqrt{2}} {v_{th}\over c} \omega_{pe} \pm {1\over \pi} \omega_{ce}\right),
\label{eq:frmWeiblim}
\end{eqnarray}
where the double sign same order.  Set the back ground magnetic field direction as positive direction of $z$ axis,
the $x,y$ components of the magnetic field of the eigen modes  
satisfy following relations 
\begin{eqnarray}
B_{kx}&=& \mp i B_{ky}, \label{eq:eigmWeiblim}
\end{eqnarray}
where the upper sign for $n=+1$ and the lower sign for $n=-1$.  
When the back ground  magnetic field is antiparallel to the temperature gradient, 
results for $n=+1$ corresponds to the above results obtained for $n=-1$ and
results for $n=-1$ corresponds to the above results obtained for $n=+1$.

\begin{figure}
\FigureFile(8cm, 8cm){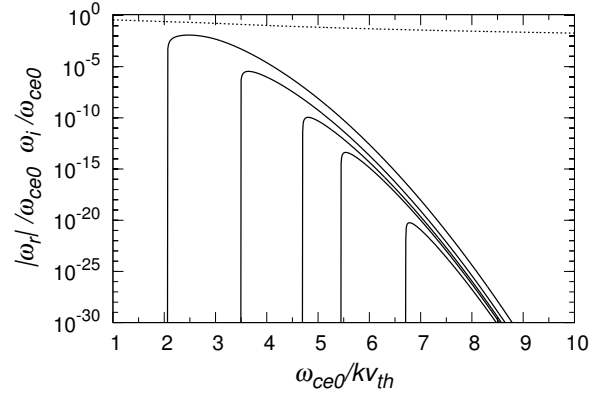}
\caption{The dispersion relations obtained by linear plasma kinetic stability analysis for a magnetised plasma.  
The equipartition magnetic field strength, that is $\beta=1$,  is assumed.  The horizontal axis is $\omega_{ce0}/k v_{th}$ 
where $k$ is absolute value of the wave vector of the mode. 
The dotted line is an absolute value of real part of frequency of unstable modes normalized by $kv_{th}$ for $\delta f_1=0.3$. 
The solid lines show imaginary parts of frequency of the unstable modes normalized by $\omega_{ce0}$ from left to right
for $\delta f_1=0.3$, 0.03, 0.01, 0.006 and 0.003, respectively.  
\label{fig:disprela}}
\end{figure}

Next the results for $|\xi| > 1$ are summarized.  
This situation corresponds to $\beta \sim 1$  or $\delta f_1 \sim 1$. 
We solved numerically the plasma dispersion function using the continued fraction expansions  (\cite{Kaji1966}, \cite{Mccabe1984})
to obtain the dispersion relation in this case since the asymptotic expansion of the plasma dispersion function can not be used. 
The plasma $\beta$ is set equal to 1 in the following analysis. 
When the direction of the temperature gradient is parallel to 
the back ground magnetic field, 
unstable modes emerge when $k_{\parallel}=\mp k$ for $n=\pm 1$ modes and
there are no unstable mode for $n=\pm 1$ when $k_{\parallel}=\pm k$ where the double sign same order. 
These results are able to be understood because the modes with $k_{\parallel} =\pm k$ for $n=\pm 1$ appeared 
in high $\beta$ plasma 
are stabilized by the back ground magnetic field as $\beta$ approaching to 1 (see Eq.(\ref{eq:omiWeiblim}))
and only $k_{\parallel}=\mp k$ for $n=\pm 1$ 
are remained unstable. 
When the direction of the temperature gradient is antiparallel to the direction of the back ground magnetic field, 
unstable modes appear when $k_{\parallel}=\pm k$ for $n=\pm 1$ where the double sign same order. 
Eigen modes of the unstable modes are circular polarized magnetic waves.
For  $n=+1$ mode when $\vec{B}_0\parallel \vec{\nabla}T_e$ and $n=-1$ when $\vec{B}_0\parallel -\vec{\nabla} T_e$, 
the eigen modes satisfy $B_{kx}=-iB_{ky}$.   
For $n=+1$ when $\vec{B}_0\parallel -\vec{\nabla}T_e$ and $n=-1$ when $\vec{B}_0\parallel \vec{\nabla} T_e$, 
the eigen modes satisfy $B_{kx}=iB_{ky}$. 
The absolute values of the real part of $\omega$ for $\delta f_1=0.3$ is shown in Fig.\ref{fig:disprela}
and is order of $\delta f_1 kv_{th}$.  
Sign of the real part is positive when $n=+1$ and is negative when $n=-1$. 
Therefore, the directions of the phase velocity of the unstable modes are parallel to the direction of the heat flow,
regardless of the direction of the temperature gradients and signs of $n$. 
This propagation direction of the wave is opposite to the propagation direction of the wave 
excited in a high $\beta$ plasma. 
The growth rates of the unstable modes have an identical shape for $n=+1$ and $n=-1$. 
The unstable modes exist only in the limited range of the wave vector around 
$\omega_{ce0}/v_{th}\sim 9\times 10^{-6}(B_0/3\mu {\rm G})/(T_e/10^6{\rm K})^{1/2}\;\;{\rm cm}^{-1}$.
The growth rates have sharp peak at the wave vector of
\begin{eqnarray}
k_m&=&{1\over a} {\omega_{ce 0}\over v_{th}}={1\over a} \left(2\over \beta\right)^{1/2} {\omega_{pe}\over c} \label{eq:kmax}
\end{eqnarray}
where $a$ runs from 2 to 5 as $\delta f_1$ runs from 0.3 to 0.002.
It is natural to assume that only the modes with $k=k_m$ are excited by the instability. 
Amplitude of the maximum growth rate decreases rapidly as  decreasing $\delta f_1$.  
When $\delta f_1$ becomes smaller than 0.002, the growth time scale  starts to exceed the age of the universe for 
typical interstellar plasma . 
So we conclude that the instability sets in only when $\delta f_1>$0.002. 
The decrease of the growth rates as decreasing $\delta f_1$ 
are much more dramatic than the decreases expected from Eq.(\ref{eq:grmWeiblim}).

The nonlinear saturation levels of the excited waves could be defined by one of the following three mechanisms 
depending on the situation. 
First argument  is that the growth of the unstable modes 
stop when the Larmor radius of thermal electron gets shorter than the wavelength of the growing mode as discussed by
 \cite{OH2003}. This is expressed as $v_{th}\omega_{ce1}^{-1}k_m\sim 1$ 
 where $\omega_{ce1}$ is electron gyro frequency with the excited wave magnetic field strength, $B_1$. 
Hereafter,  this is referred to thermal electron trapping condition. 
By taking into account the possibility that the growth of the wave stops before it gets 
$v_{th}\omega_{ce1}^{-1}k_m\sim 1$, 
the saturation level of the magnetic field strength of the unstable modes is expressed as 
\begin{eqnarray} 
\omega_{ce 1}&=& b k_m v_{th}, \label{eq:omece1}
\end{eqnarray}
where  $b$ is a non dimensional number satisfying $b\le 1$. 
Second argument is that the saturation level is determined by balancing between growth and nonlinear 
damping of the waves (\cite{RL1978}, \cite{LE1992}). 
\cite{LE1992} estimated  the nonlinear damping rate due to wave-wave interaction 
for magnetized plasma as $\gamma^{nl}\sim v_{th}/(\lambda_{eff} \beta(1+\omega_{ce0}/k_mv_{th})^{2})
 \sim v_{th}\delta_T/(\tilde{\epsilon}\delta_T \delta x\beta(1+\omega_{ce0}/k_mv_{th})^{2})$
 where $\lambda_{eff}$ is an effective electron mean free path reduced from the Coulomb collision mean free path
 due to the scattering by the excited magnetic waves and $\tilde{\epsilon}$ appeared in $\gamma_{nl}$ is defined 
 by the ratio between the effective mean free path 
 to the scale of the temperature gradient, $\delta x$.
 The  $\tilde{\epsilon}$ may decrease from the initial value of $\epsilon$ because the effective mean free path becomes shorter as
 a fraction of trapped thermal electrons which are trapped by the magnetic field of the excited waves, is increased 
 as the amplitude of the magnetic field of the excited wave is growing.  
 The decrease may stop at the value at where 
the growth rate balances with the decay rate.
Since the quantitative arguments how the non linear saturation level is defined by this condition 
depend on detail models of the temperature gradient of the system, 
we skip further quantitative arguments based on this argument. 
Third argument 
is whether initial energy content is enough 
to excite the magnetic plasma waves.
\cite{Kato2005} showed by performing plasma particle simulations aiming to study the nonlinear evolution 
of the Weibel instability that 
the initial difference of the kinetic energy caused by the anisotropy of the temperature of the electrons,
that is $\Delta W\sim k_B T_1-k_B T_2$ where $T_1$ and $T_2$ are 
temperatures in two different directions, 
is converted into the magnetic energy of the excited waves when the amplitude of the waves get maximum value. 
As pointed out by \cite{OH2003}, the physical essence of the instability driven by the temperature gradient is 
equivalent to the Weibel instability with a directional temperature difference of 
$\Delta T\sim (\delta f_1)^2T_e$. 
Therefore, the maximum available energy stored in the magnetic field of the excited waves 
is about $k_B (\delta f_1)^2T_e$.
This sets upper limit on the excited magnetic field strength as 
\begin{eqnarray}
2n_e k_B \delta f_1^2 T_e&>& {B_1^2\over 8\pi}. \label{eq:nsenearg}
\end{eqnarray}
For high $\beta$ plasma, this condition sets upper limit on $b$ appeared in Eq.(\ref{eq:omece1}) as
\begin{eqnarray}
b&\le& \beta^{1/2} a \delta f_1. \label{eq:buplim}
\end{eqnarray}
In the case of high $\beta$ plasma, 
 as long as $b<1$ the condition (\ref{eq:nsenearg}) is satisfied.

\section{Theory of the Jitter radiation in a magnetized plasma}

\subsection{Fundamentals}

The emission mechanisms when 
relativistic electrons travel through the magnetized plasma which is filled with the 
circular polarized magnetic waves excited by the RL instability are examined. 
When the coherent length of the magnetic fields is much longer than the gyration radius of the relativistic electrons,  
hereafter we refer this field ordered field,
the emitted radiation is known as synchrotron radiation. 
In this section, we drive the radiation spectrum when relativistic electrons travel through the plasma 
which is occupied by both ordered magnetic fields and  magnetic waves generated by the 
RL instability discussed in Section 2. 
There are similarities in procedures with the derivation of the weak undulator radiation spectrum (\cite{Hof2004}).
Set the direction of the ordered magnetic field to $z$-axis as
$\vec{B}_0=(0,0,B_0)$. 
Assume that the gradient of the electron temperature is parallel to the ordered magnetic field direction. 
Consider unstable mode with $n=+1$ in the following analysis. 
As discussed in Sec.2, the unstable mode is limited for the waves  with $k_z<0$ where $k_z$ is a $z$ component  of the wave vector of the waves in the case of the low $\beta$ plasma.
Since the growth rate takes the maximum value at $k=k_m$,
we assume that  the excited waves due to the instability is 
monochromatic with the wave vector of $\vec{k}=(0,0,-k_m)$.   
Hereafter, we refer $k_m=k$.

The zeroth order motion of relativistic electrons are given by 
following equations of motion 
\begin{eqnarray*}
\gamma m_e{d v_{x0} \over d t}&=& -{e\over c} v_{y0} B_0,\\
\gamma m_e{d v_{y0}\over dt} &=& {e\over c} v_{x0} B_0,\\
{d v_{z0}\over dt} &=&0,
\end{eqnarray*}
where $\gamma$ is a Lorentz factor of the relativistic electron. 
The zeroth order orbits for relativistic electron with a velocity of $v_0$  (an equivalent Lorentz factor of $\gamma$) and a
pitch angle of $\alpha$ are given by solving these equations as
\begin{eqnarray*}
v_{x0}= v_{\perp 0} \cos \omega_{se0} t,&& x_0={v_{\perp 0}\over \omega_{se 0}} \sin \omega_{se 0} t,\\
v_{y0}=v_{\perp 0}\sin \omega_{se0} t,&& y_0=-{v_{\perp 0} \over \omega_{se 0} }\cos \omega_{se 0} t,\\
v_{z0}= v_{\parallel 0},&& z_0=v_{\parallel 0} t,
\end{eqnarray*}
where  $v_{\perp 0}\equiv v_0 \sin \alpha,\; v_{\parallel 0}\equiv v_0 \cos \alpha$ and $\omega_{se 0} \equiv \omega_{ce 0}/\gamma$. 
The zeroth order orbit is helical motion around the ordered magnetic field with a gyration period of $T=2\pi/\omega_{se 0}$. 

Orbits of the relativistic electrons are perturbed due to 
the magnetic waves excited by the instability.
The similar studies in the  context of scattering of relativistic charged particles
 by magnetic irregularities was studied by \cite{Parker1964}. 
The magnetic field of the unstable mode for $n=+1$ is expressed as 
\begin{eqnarray*}
B_{x1}&=&B_1{\cos} (k_mz+\omega_r t),\\
B_{y1}&=&B_1{\sin} (k_mz+\omega_r t),
\end{eqnarray*}
where $\omega_r$ is a real part of the angular frequency of the unstable mode.
Velocities and orbits generated by the wave fields are described by 
$\vec{v}_1$ and  $\vec{x}_1$. 
In the first order of $v_1/v_0$ and $B_1/B_0$, the equations of motion for $\vec{v}_1$ are given by
\begin{eqnarray*}
\gamma m_e {d v_{x1}\over dt}&=& -{e\over c} v_{y1} B_0 + {e\over c} v_{z0} B_{y1},\\
\gamma m_e {d v_{y1}\over dt} &=& {e\over c} v_{x1} B_0 -{e\over c} v_{z0} B_{x1},\\
\gamma m_e {d v_{z1}\over dt} &=& -{e\over c} v_{x0} B_{y1} +{e\over c} v_{y0} B_{x1}.
\end{eqnarray*}
By inserting the zeroth order orbit as for the orbit of electron, the wave fields are 
approximately described as 
\begin{eqnarray*}
B_{x1}&=&B_1 \cos (k v_{\parallel 0} +\omega_r) t,\\
B_{y1}&=&B_1 \sin (k v_{\parallel 0} +\omega_r) t.
\end{eqnarray*}
The perturbed velocities in the first order are obtained as 
\begin{eqnarray}
v_{x1}+iv_{y1}&=&{-\omega_{se1} v_{\parallel 0}\over kv_{\parallel 0} +\omega_r-\omega_{se 0}} e^{i(kv_{\parallel 0} +\omega_r)t} 
+C e^{i\omega_{se 0} t} \label{eq:vxy1},\\
v_{z1}&=& {\omega_{se 1} v_{\perp 0}\over k v_{\parallel 0} +\omega_r-\omega_{se 0} } \cos(k v_{\parallel 0} +\omega_r -\omega_{se 0} )t.\label{eq:vz1}
\end{eqnarray}
Since the last term in Eq.(\ref{eq:vxy1}) is describing the gyration motion around the ordered field, 
it is able to be absorbed in the zeroth order motion and we set $C=0$.
For high $\beta$ plasma 
\begin{eqnarray*}
{\omega_r \over k v_{\parallel 0}}&\sim& {v_{th}\over c} {\omega_r \over k v_{th}}\ll 1,
\end{eqnarray*}
and for low $\beta$ plasma 
\begin{eqnarray*}
{\omega_r \over k v_{\parallel 0}}&\sim& a\delta f_1 {v_{th}\over c} \ll 1.
\end{eqnarray*}
Further,
\begin{eqnarray*}
{\omega_{se 0} \over k v_{th}}&\sim & {a\over \gamma} {v_{th}\over c} \ll 1,
\end{eqnarray*}
is satisfied. 
Therefore, 
amplitudes in Eqs. (\ref{eq:vxy1}) and (\ref{eq:vz1}) are approximately $\omega_{se 1}/k$. 
As discussed in below, the contribution of the perturbed motion of the relativistic electron to the radiation spectrum 
happens within a only limited time interval of $T'\sim 1/\omega_{ce 0}$.
Changes of the phases in Eqs.  (\ref{eq:vxy1}) and (\ref{eq:vz1})  within this time interval are 
$k v_{\parallel 0} T' \gg 1$,  $\omega_{se 0} T'\sim {1\over \gamma}\ll 1$ and  $\omega_r T' \sim \delta f_1< 1$.
Therefore, the phase factors in Eqs. (\ref{eq:vxy1}) and (\ref{eq:vz1}) are approximated as $k v_{\parallel 0} t$ 
in the following discussion. 
Under these approximations, the velocities and the orbits of the perturbed motion are described by 
\begin{eqnarray}
v_{1x}&=&-{\omega_{se 1}\over k} \cos kv_{0\parallel} t,\label{eq:vx1d}\\
v_{1y}&=&-{\omega_{se 1}\over k} \sin kv_{0\parallel} t,\label{eq:vy1d}\\
v_{1z}&=&{\omega_{se 1}\over k} \tan\alpha\cos kv_{0\parallel}t, \label{eq:vz1d}\\
x_1&=& -{\omega_{se 1}\over k} {1\over k v_{\parallel 0}} \sin k v_{\parallel 0} t,\label{eq:x1}\\
y_1&=& {\omega_{se 1}\over k} {1\over k v_{\parallel 0}} \cos k v_{\parallel 0} t,\label{eq:y1}\\
z_1&=& {\omega_{se 1}\over k} {1\over k v_{\parallel 0}}\tan \alpha \sin k v_{\parallel 0} t.\label{eq:z1}
\end{eqnarray}
Relative amplitudes of the velocity and the position of the perturbed motion to the zeroth order orbit are  
\begin{eqnarray}
\left| {v_{\perp 1}\over v_{\perp 0}}\right| \sim \left| {v_{z1}\over v_{z0}}\right| &\sim&  {ab\over \gamma} {v_{th}\over c} \ll 1,\label{eq:compv}\\
\left| {r_{\perp 1}\over r_{\perp 0} }\right| \sim \left|{z_1\over z_0}\right| &\sim& {ab\over \gamma^2} \left({v_{th}\over c}\right)^2 \ll 1, \label{eq:compr}
\end{eqnarray} 
for low $\beta$ plasma 
where we implicitly assume that the pitch angle does not take a value  close to $\pi/2$ so as to avoid being $\tan\alpha\gg 1$,
and $v_{\perp 1}=\sqrt{v_{x1}^2+v_{y1}^2}$ and $r_{\perp 1}=\sqrt{x_1^2+y_1^2}$.
For high $\beta$ plasma, 
\begin{eqnarray}
\left| {v_{\perp 1}\over v_{\perp 0}}\right| \sim \left| {v_{z1}\over v_{z0}}\right| &\sim&  {1\over \gamma} {v_{th}\over c} \beta^{-1/2}{1\over \delta f_1},\label{eq:compvHB}\\
\left| {r_{\perp 1}\over r_{\perp 0} }\right| \sim \left|{z_1\over z_0}\right| &\sim& {1\over \gamma^2} \left({v_{th}\over c}\right)^2 
\beta^{-1} \delta f_1^{-2}. \label{eq:comprHB}
\end{eqnarray} 
Therefore, unless the amplitude of the anisotropic velocity distribution is close to or less than $v_{th}/(\gamma c)\beta^{-1/2}$, 
Eqs.(\ref{eq:compvHB}) and (\ref{eq:comprHB}) are much smaller than 1. 
These results guarantee the validity of our perturbative treatment for describing the motion of the relativistic electron. 
When the pitch angle closes to $\pi/2$, 
the amplitude of $v_{1z}$ becomes large and the perturbative treatment is broken. 
By comparing (\ref{eq:vz1d}) and the $z$ component of the zeroth order electron velocity, 
the range of the pitch angle for which the perturbative treatment is acceptable, is obtained as 
\begin{eqnarray}
\alpha&<& {\pi\over 2} -\sqrt{{1\over \gamma} {v_{th}\over c}}. \label{eq:condalpha}
\end{eqnarray}
As far as the pitch angle is in this range, the amplitude of the perturbed velocities are less than the amplitude of 
the zeroth order velocities. 
Therefore, except a limited small range of the pitch angle, our treatment is valid.

The Fourier spectrum of the electric field of the radiation emitted from a single relativistic electron is obtained by inserting the electron orbit in the 
following equation 
\begin{eqnarray}
R\vec{\hat{E}}(\omega)&=&{i e\omega \over 2\pi c} e^{i\Phi}\int^{T_2'}_{T_1'}(\vec{n}\times (\vec{n}\times \vec{\beta}_0))e^{i\omega \left(t'-{1\over c} \vec{n}\cdot\vec{r}(t')\right)}
dt'  \nonumber\\
&+& {i e\omega \over 2\pi c} e^{i\Phi}\int^{T_2'}_{T_1'}(\vec{n}\times (\vec{n}\times \vec{\beta}_1))e^{i\omega \left(t'-{1\over c} \vec{n}\cdot\vec{r}(t')\right)}
dt' ,
\label{eq:Espec0}
\end{eqnarray}
where $\vec{\beta}_0=\vec{v}_0/c$ and $\vec{\beta}_1=\vec{v}_1/c$, and $\vec{r}=\vec{r}_0+\vec{r}_1$ is superposition of 
a zeroth order orbit and an orbit of the  perturbed motion of the electron.  
The first term contributes the synchrotron radiation. 
The contribution to the radiation spectrum of the  perturbed orbit, $\vec{r}_1$, appeared in the exponent of the first term is negligible. 
The second term provides the Jitter radiation. 
Similar analysis was done by \cite{GS1965}.
They studied the polarization degree of the synchrotron emission when a random field is
superimposed on a homogeneous field.

\begin{figure}
\FigureFile(8cm, 8cm){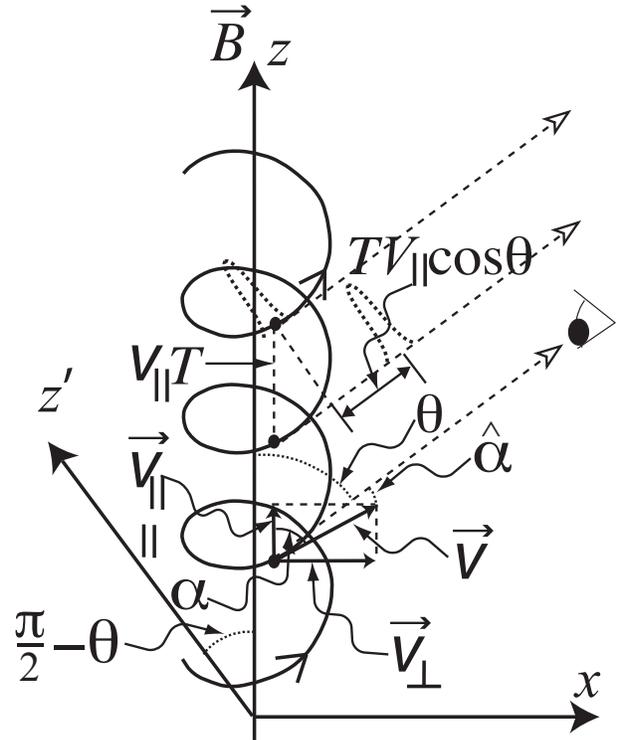}
\caption{Orbit of a relativistic electron under back ground magnetic field. The electron travels helical orbit. 
The line of sight is taken in $x$-$z$ plane for simplicity and is described by dashed line with arrow. 
Inclination angle of the line of sight relative to $z$ axis is defined as $\theta$.  By rotating the coordinate with angle $\theta$ 
as an axis in $y$-axis, $x$-$z$ is transformed into $x'$-$z'$.  The $x'$ axis coincide with the line of sight.  The pitch angle of the
electron is defined as $\alpha$. The deviation of the pitch angle from $\theta$ is denoted by $\hat{\alpha}$.  
\label{fig:rasen}}
\end{figure}

\begin{figure}
\FigureFile(8cm, 8cm){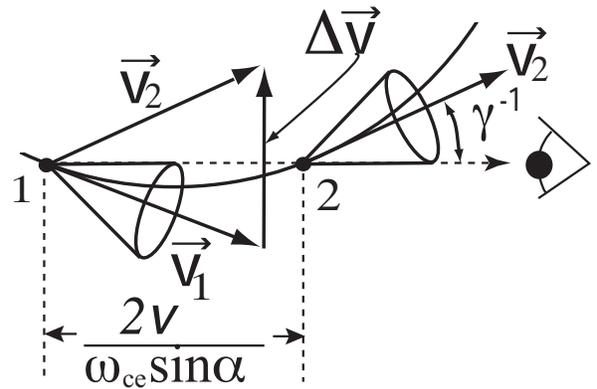}
\caption{Part of a relativistic electron orbit in $x'$-$y$ plane for an electron with $\alpha=\theta$. 
The emission cones due to the relativistic beaming effect are described by cones with  
a solid angle of $\pi/\gamma^2$ assigned to electrons which are denoted by black dots.
\label{fig:toudai}}
\end{figure}

\begin{figure}
\FigureFile(8cm, 8cm){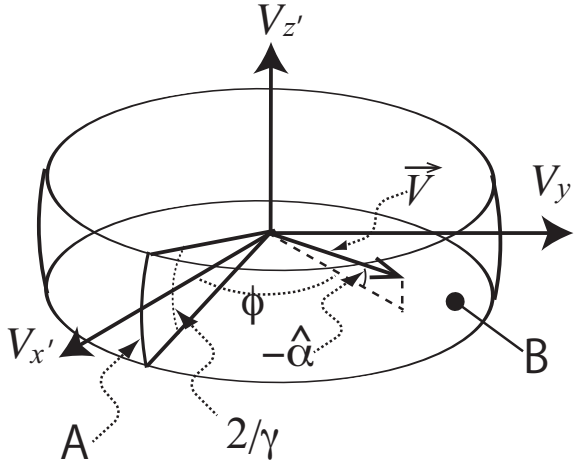}
\caption{The zeroth order velocity distribution of the relativistic electrons with a fixed $\gamma$ 
which contribute to the Jitter radiation are described in velocity space.
The $x'$, $y$ and $z'$ components of velocities are denoted by $v_{x'}$, $v_{y}$ and $v_{z'}$, respectively. 
An arc labeled by A denotes the velocities spanned by $-1/\gamma\le \hat{\alpha} \le 1/\gamma$
where $\hat{\alpha}$ is relative pitch angle of electrons to the line of sight. 
The arc A is a part of a circle in $v_{x'}$-$v_{z'}$ plane with a radius of $v$ centered on the origin. 
The area labeled by B is made by rotating the arc A around $v_{z'}$ axis. 
This surface is a part of the sphere with a radius of $v$ centered on the origin. 
The electrons with velocities which point to the area B in velocity space 
at some fixed moment contribute to the Jitter radiation once every period of $T$.   \label{fig:radicone}}
\end{figure}

The radiation spectrum contributed from the second term of equation (\ref{eq:Espec0}) is evaluated as follows. 
The direction cosine of the line of sight is defined in $x$-$z$ plane as $\vec{n}=(\sin \theta, 0, \cos\theta)$.
When the motion of the electron is relativistic, the radiation from the electron is concentrated in the 
so called emission cone which is the cone with solid angle of $\pi/\gamma^2$
due to the relativistic beaming effect.
Therefore, the radiation from electrons are observed only when the line of sight is included in the emission cones.   
In Figure \ref{fig:rasen}, the zeroth order orbit of an electron is described.  
It shows that electron travels along a helical orbit.
The line of sight is described by a dashed line with arrow. 
In Figure \ref{fig:toudai}, a part of the electron orbit in $x'$-$y$ plane when $\alpha=\theta$ is described. 
When the electron arrives at the position labeled by 1, 
the emission cone starts to include the line of sight. 
When the electron reaches at the position labeled by 2, 
the line of sight exits the emission cone. 
During this interval, the line of sight is kept to be included in the emission cone. 
Hereafter we refer this interval as emission orbit. 
A duration while an electron travels the emission orbit is $T'=2/(\omega_{ce} \sin\theta)$.
Since the small deviation of the pitch angle of the electron from the line of sight is important to determine the characteristics of the Jitter radiation,
the pitch angle is expressed as $\alpha=\theta+\hat{\alpha}$ in the following discussion as shown in Figure \ref{fig:rasen}.  
In Figure \ref{fig:radicone}, 
the zeroth order velocity distribution of the electrons with a fixed $\gamma$ 
which contribute to the Jitter radiation is  shown. 
In this Figure, $v_{x'}$ axis is taken to the direction of the line-of-sight 
and $v_{z'}$ axis is taken to the direction normal to the plane spanned by $v_{x'}$ and $v_y$ axes as shown in Figure \ref{fig:rasen}. 
The relative pitch angle to the line of sight, $\hat{\alpha}$, is shown in this figure.  
In the following discussion, the amplitude of the velocity, e.g.  $\gamma$, is fixed for simplicity. 
An arc labeled by A denotes the velocities spanned by $-1/\gamma\le \hat{\alpha} \le 1/\gamma$. 
The arc A is a part of a circle in $v_{x'}$-$v_{z'}$ plane with a radius of $v$ centered on the origin. 
The area labeled by B is made by rotating the arc A around $v_{z'}$ axis. 
Therefore, this surface is a part of the sphere with a radius of $v$ centered on the origin. 
Since an electron velocity is contained in $v_{x'}$-$v_{z'}$ plane every $T$ as illustrated in Figure \ref{fig:rasen},
electrons with velocities which direct to the area B at some fixed moment contribute to the Jitter radiation once every $T$.   

\subsection{Essence of the Jitter radiation}

Firstly, the physical characteristics of the Jitter radiation in the low $\beta$ plasma are sketched. 
By performing the coordinate transformation from $v_x$-$v_z$ to $v_{x'}$-$v_{z'}$, the first order perturbed velocities are transformed as
\begin{eqnarray}
v_{x'}&=&{\omega_{se 1}\over k}{\hat{\alpha} \over \cos\theta}  \cos k v_{0 \parallel} t',\label{eq:vx1p}\\
v_y&=&-{\omega_{se 1}\over k}\sin k v_{0 \parallel} t',\label{eq:vy1p}\\
v_{z'}&=& {\omega_{se 1}\over k} {1\over \cos\theta} \cos k v_{0 \parallel} t',\label{eq:vz1p}
\end{eqnarray}
where the results are expanded in the lowest order of $\hat{\alpha}$ and $t'$ is used to describe  
time at the position of  the electron. 
The trajectory of the electron total velocity vector in velocity space
is described in Figure \ref{fig:prec}. 
The total velocity is a sum of the zeroth order velocity and the first order velocity.
The velocity vectors  of the electrons precess around the line of sight. 
 Direction of the rotation is clock wise rotation from the observer. 
 The velocity vector traces elliptical trajectory with a minor radius of $\omega_{se1}/k$ and a major radius of $\omega_{se1}/(k\cos\theta)$.
 In other word, the directions of the motion of electrons precess around the line of sight while changing inclination angle from $\omega_{se 1}/kc$ to
 $\omega_{se 1}/(kc \cos\theta)$ as shown in Figure \ref{fig:prec}. 
 As far as the line of sight is not close to perpendicular direction to the back ground magnetic field and $\hat{\alpha}<1/\gamma$,  
 the line of sight is always included in the emission cones of precessing electrons since
  \begin{eqnarray}
{ \omega_{se 1}\over k c}&\sim& {b\over \gamma} {v_{th}\over c} \ll {1\over \gamma}. \label{eq:incangl}
\end{eqnarray}
The period of the precession is much shorter than the duration while electrons travel the emission orbit since
\begin{eqnarray}
{2/\omega_{ce 0} \sin\theta\over 2\pi/kc\cos\theta}&=& {1\over \pi a} {c\over v_{th}}\cot\theta \gg 1. \label{eq:prec2deo}
\end{eqnarray}
Therefore, the electron emits the radiation at the frequency 
of the inverse of the observed period of the precession of the electron.
This is the Jitter radiation.  
This situation is illustrated in Figure \ref{fig:jitter}. 
Suppose that an electron emits electromagnetic wave when it arrives at the position 
labeled by 1 at the time $t_1'$.  
The distance between the position 1 and the observer is $L$. 
The electromagnetic wave emitted at the position 1 arrives at the observer at $t_1=t_1'+L/c$. 
During one period of the precession, the electron moves from position 1 to position 2.
The distance traveled by the electron during the period  
is $2\pi/(k\cos\theta)$ which is an effective wave length of the excited plasma wave
for the observer whose line of sight is inclined by an angle $\theta$  relative to the wave vector of the plasma wave. 
The electron reaches at position 2 at $t_2'=t_1'+2\pi/(k\cos\theta\; v_0)$. 
The arrival time of the electromagnetic wave emitted at position 2 to the observer is 
$t_2=t_2'+(L-2\pi/(k \cos\theta))/c$. 
During this period, the electron emits right handed elliptically polarized electromagnetic waves toward the observer
where the polarization pattern traces the orbit of the electron illustrated in Figure \ref{fig:prec}.   
Since the line of sight is always included in the emission cone during this time interval, 
all the emission emitted in this time interval is observed by the observer
except the case that the direction of the magnetic field is close to perpendicular to the line of sight. 
The observed period of the electromagnetic wave, $\Delta t=t_2-t_1$ is given by
\begin{eqnarray}
\Delta t&\sim&{2\pi\over kc\cos\theta} \left(1-{v_0\over c}\right)\sim {\pi\over \gamma^2 kc \cos\theta}. \label{eq:Jitperiod}
\end{eqnarray}
Therefore, the observed frequency of the Jitter radiation is 
\begin{eqnarray}
\nu_{Jit}&\sim&{\cos\theta\over a\pi} \sqrt{2} \beta^{-1/2} \gamma^2 \omega_{pe},\label{eq:Jitfreq}
\end{eqnarray}
where Eq.(\ref{eq:kmax}) was used. The emitted frequency by the Jitter radiation is the Doppler shifted plasma frequency in $\beta\sim 1$
plasma. 
On the other hand, the frequency of the synchrotron emission(\cite{Rybicki1979}) is given by
\begin{eqnarray}
\nu_{syn}&\sim& 0.29 {1\over 2\pi} {3\over 2} \gamma^2 \omega_{ce0} \sin\theta. \label{eq:synfreq}
\end{eqnarray}
The ratio of these two frequencies for a fixed $\gamma$ is written as
\begin{eqnarray}
{\nu_{Jit}\over \nu_{syn}}&\sim& {4 \over a} \cot\theta \left({v_{th}\over c}\right)^{-1}= {4\over a} \cot\theta \sqrt{m_e c^2\over 2k_B T}. \label{eq:Jit2syn}
\end{eqnarray} 
It shows that as far as electron thermal velocity is non relativistic,  an relativistic electron with a fixed $\gamma$ 
emits much higher frequency by the Jitter radiation compared with 
the synchrotron emission.  
The condition (\ref{eq:prec2deo}) must be satisfied for the Jitter radiation to be observed.  
As the line of sight approaching to the perpendicular direction of the magnetic field, 
the period of the precession approaches to the duration while electrons travel the emission orbit. 
When the line of sight is in the range of
\begin{eqnarray}
{\pi\over 2}&\ge& \theta \ge {\pi\over 2} -a\pi {v_{th}\over c}, \label{eq:condNoJit}
\end{eqnarray}
the precession period becomes longer than the duration while electrons travel the emission orbit. 
It shows that except the small limited range of $\theta$ the Jitter radiation is observable in any line of sight relative to the 
magnetic field direction. 

\begin{figure}
\FigureFile(8cm,8cm){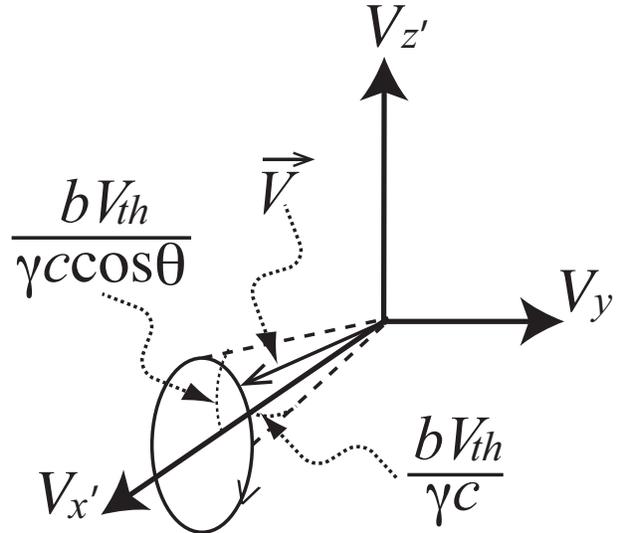}
\caption{A trajectory of velocity vector of a relativistic electron  in velocity space while the electron travels 
emission orbit. The velocity vector precesses around $v_{x'}$ axis. 
It rotates clock wise direction from the observer along elliptical trajectory with a minor radius of $\omega_{se1}/k$ and 
a major radius of $\omega_{se1}/(k\cos\theta)$ denoted by thick solid line.
\label{fig:prec}}
\end{figure}

\begin{figure}
\FigureFile(8cm,4cm){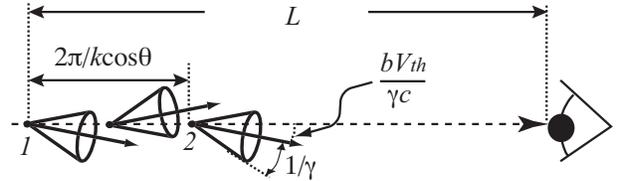}
\caption{Properties of jittering motion of relativistic electron while it travels emission orbit. The electron approaches 
to the observer while it carries out the precession described in Fig.\ref{fig:prec}. The line of sight is 
always included in the emission cone while the electron travels the emission orbit. 
During one period of the precession, the electron moves a distance of $2\pi/(k\cos\theta)$. \label{fig:jitter}}
\end{figure}

\subsection{Analytical formulation of the spectrum of the Jitter radiation}

In this subsection, analytical formulae of the spectrum of the Jitter radiation are provided.  
First the spectrum emitted by  a single electron is evaluated. 
Basic equations are the second term of right hand side of Eq.(\ref{eq:Espec0}).
Since the Jitter radiation is emitted only when the electrons travel the emission orbit, 
we adopt following approximations. 
When the electron is traveling the emission orbit, 
the zeroth order orbit is approximated as straight line.
Then, a part of the phase appeared in Eq.(\ref{eq:Espec0}) is evaluated as 
\begin{eqnarray}
\omega \left(t'-{1\over c} \vec{n}\cdot\vec{r}_0\right)&\sim& {\omega \over \kappa(\hat{\alpha})}t',\label{eq:exp0}\\
\kappa(\hat{\alpha})&\equiv& {2\gamma^2\over 1+\gamma^2\hat{\alpha}^2},\label{eq:kappa}
\end{eqnarray}
when $-T'/2\le t' \le T'/2$.
When the electron is out of the emission orbit, that is when $-T/2\le t' < -T'/2$ or $T'/2< t' \le T/2$, 
we assume that there is no contribution from second term.  
In the following calculation, we use $x'$-$y$-$z'$ coordinate instead of $x$-$y$-$z$. 
In $x'$-$y$-$z'$, the line of sight is $\vec{n}=(1,0,0)$. 
Then the Fourier spectrum of each component of the 
electric field of the emitted radiation are deduced as 
\begin{eqnarray}
R\hat{E}_{x'}(\omega)&=&0\label{eq:Especx},\\
R\hat{E}_{y}(\omega)&=& {ie\omega \over 2\pi c} e^{i\Phi}\int_{-{T'\over 2}}^{T'\over 2} dt'{\omega_{se1}\over k c}\sin k v_{0\parallel}t'\nonumber\\
&&\times
e^{i\omega\left({t'\over \kappa(\hat{\alpha})}-{\omega_{se 1}\over k^2 c v_{0\parallel}}{\hat{\alpha}\over \cos\theta} \sin k v_{0\parallel} t'\right)}\label{eq:Especy0},\\
R\hat{E}_{z'}(\omega)&=&-{ie\omega \over 2\pi c} e^{i\Phi}\int_{-{T'\over 2}}^{T'\over 2} dt'{\omega_{se1}\over k c\cos\theta}\cos k v_{0\parallel}t'\nonumber\\
&&\times e^{i\omega\left({t'\over \kappa(\hat{\alpha})}-{\omega_{se 1}\over k^2 c v_{0\parallel}}{\hat{\alpha}\over \cos\theta} \sin k v_{0\parallel} t'\right)}.\label{eq:Especz0}
\end{eqnarray}
By using following formula for the Bessel functions
\begin{eqnarray}
e^{-i\lambda \sin k v_{0\parallel} t'}&=&\sum_{n=-\infty}^{\infty} J_n(\lambda) e^{-inkv_{0\parallel} t'},\label{eq:Bessel}
\end{eqnarray}
equations (\ref{eq:Especy0}) and (\ref{eq:Especz0}) are evaluated as 
\begin{eqnarray}
R\hat{E}_y(\omega)&=&{e\omega \omega_{se1}\over 4\pi k c^2}e^{i\Phi}\sum_n(J_{n+1}(\lambda)-J_{n-1}(\lambda))T'\nonumber\\
&&\times {\rm sinc}\left({\omega\over \kappa(\hat{\alpha})}-nkv_{0\parallel}\right){T'\over 2},\label{eq:Especy1}\\
R\hat{E}_{z'}(\omega)&=&-{ie\omega \omega_{se1}\over 4\pi k c^2\cos\theta }e^{i\Phi}\sum_n(J_{n+1}(\lambda)+J_{n-1}(\lambda))T'\nonumber\\
&&\times{\rm sinc}\left({\omega\over \kappa(\hat{\alpha})}-nkv_{0\parallel}\right){T'\over 2},\label{eq:Especz1}
\end{eqnarray}
where $\lambda$ is dimensionless parameter defined as 
\begin{eqnarray}
\lambda&\equiv& {\omega \omega_{se 1}\over k^2 c v_{0\parallel}\cos\theta}\hat{\alpha}.\label{eq:lambda}
\end{eqnarray}
Since the sinc function appeared in these equations plays a role  such as a delta function,
the angular frequency $\omega$ is replaced by $\omega_n=nkv_{0\parallel} \kappa(\hat{\alpha})$. 
Then the order of magnitude of $\lambda$ for each $\omega_n$ is estimated with a help of Eq.(\ref{eq:omece1}) as 
\begin{eqnarray}
\lambda_n&=&{\omega_n\over k v_{0\parallel}} {\omega_{se1}\over kc} {\hat{\alpha}\over \cos\theta} \sim {2nb\over \cos\theta}\gamma {v_{th}\over c}\hat{\alpha}
< {2nb\over \cos\theta} {v_{th}\over c}\ll n, \label{eq:lambdan}
\end{eqnarray}
where $|\hat{\alpha}|<1/\gamma$ is used since the main contribution comes from the electrons with the pitch angle $|\hat{\alpha}|<1/\gamma$.
Since $J_0 (0)=1$ and $J_n(\lambda_n)\ll 1$ for $\lambda_n\ll n$ when $n\neq 0$, 
only the terms which contain $J_0$ provide main contributions in Eqs.(\ref{eq:Especy1}) and (\ref{eq:Especz1}). 
Since we are only  interested in positive frequency,  we obtain 
\begin{eqnarray}
R\hat{E}_y(\omega)&=&-{e\omega \omega_{se 1}\over 4\pi k c^2} e^{i\Phi}T'{\rm sinc}  
\left({\omega\over \kappa(\hat{\alpha})}-kv_{0\parallel}\right){T'\over 2},\nonumber\\
&&\;\;\;\;\;\;\;\;\;\;\;\;\;\;\;\;\;\;\;\;\;\;\;\;\;\;\;\;\;\;\;\;\;\;\;\;\;\;\;\;\;\;\;\;\;\;\;\;\;\;\;{\rm for}\;\;\;\theta\le {\pi\over 2} \nonumber\\
&=&{e\omega \omega_{se 1}\over 4\pi k c^2} e^{i\Phi}T'{\rm sinc}  
\left({\omega\over \kappa(\hat{\alpha})}+kv_{0\parallel}\right){T'\over 2},\nonumber\\
&&\;\;\;\;\;\;\;\;\;\;\;\;\;\;\;\;\;\;\;\;\;\;\;\;\;\;\;\;\;\;\;\;\;\;\;\;\;\;\;\;\;\;\;\;\;\;\;\;\;\;\;{\rm for}\;\;\;{\pi\over 2} < \theta\le\pi \label{eq:Especy2}\\
R\hat{E}_{z'}(\omega)&=&-{ie\omega \omega_{se1}\over 4\pi k c^2\cos\theta }e^{i\Phi}T'
{\rm sinc}\left({\omega\over \kappa(\hat{\alpha})}-kv_{0\parallel}\right){T'\over 2},\nonumber\\
&&\;\;\;\;\;\;\;\;\;\;\;\;\;\;\;\;\;\;\;\;\;\;\;\;\;\;\;\;\;\;\;\;\;\;\;\;\;\;\;\;\;\;\;\;\;\;\;\;\;\;\;{\rm for}\;\;\;\theta\le {\pi\over 2} \nonumber\\
&=&-{ie\omega \omega_{se1}\over 4\pi k c^2\cos\theta }e^{i\Phi}T'
{\rm sinc}\left({\omega\over \kappa(\hat{\alpha})}+kv_{0\parallel}\right){T'\over 2}.\nonumber\\
&&\;\;\;\;\;\;\;\;\;\;\;\;\;\;\;\;\;\;\;\;\;\;\;\;\;\;\;\;\;\;\;\;\;\;\;\;\;\;\;\;\;\;\;\;\;\;\;\;\;\;\;{\rm for}\;\;\;{\pi\over 2}<\theta\le \pi\label{eq:Especz2}
\end{eqnarray}
It shows that the emitted Jitter radiation is right handed elliptically polarized light which has a major axis in $z'$ and minor axis in $y$, and the 
ratio of the major to the minor radius is $1/\cos\theta$. 

The emitted power of the Jitter radiation per unit solid angle in the direction of $\vec{n}$ is obtained as 
\begin{eqnarray}
{d p^e_{Jit}(\omega_1)\over d \Omega}&=&c\int_0^{\infty}d\omega {|R\hat{E}_{x'}|^2+|R\hat{E}_y|^2+|R\hat{E}_{z'}|^2\over T},\nonumber\\
&=&{T'\over T} {e^2\omega_{se1}^2 \omega_1^2\over 8 \pi k^2 c^3}(2+\tan^2\theta) \kappa(\hat{\alpha}), \label{eq:pJit0}\\
\omega_1&\equiv& \kappa(\hat{\alpha}) k v_{0\parallel}={\omega_J\over 1+\gamma^2\hat{\alpha}^2}, \label{eq:ome1}\\
\omega_J&\equiv& 2\gamma^2 \omega_0, \label{eq:omeJ}\\
\omega_0&\equiv& |k v_{0\parallel}|. \label{eq:ome0}
\end{eqnarray}
As the line of sight approaching to the direction of the magnetic field, an interval of the arrival of pulses to the observer 
emitted by a single electron 
while it is in the emission orbit, becomes shorter. 
When $\theta <1/\gamma$,  the line of sight is  continuously included in the emission cone of the single electron. 
Therefore, $T'/T$ must be set 1 in Eq.(\ref{eq:pJit0})  when $\theta <1/\gamma$.
Since the length of the emission orbit becomes shorter as increasing $|\hat{\alpha}|$ by a factor of $\sqrt{1-\gamma^2\hat{\alpha}^2}$, 
the duration while a single electron is traveling the emission orbit, $T'$, is reduced to
$T'=(2/\omega_{ce}\sin\theta)\sqrt{1-\gamma^2\hat{\alpha}^2}$.  
The spectrum of the Jitter radiation is monochromatic and almost delta function centered on $\omega=\omega_1$. 
Suppose that the relativistic electron has an isotropic velocity distribution and its energy distribution is given by 
the following power law spectrum 
\begin{eqnarray}
N_e(\gamma)&=&C_p\gamma^{-p},\;\;\;\;\;{\rm for}\;\;\;\;\gamma_1\le \gamma \le \gamma_c\label{eq:Nep}
\end{eqnarray}
where $N_e(\gamma)d\gamma$ gives electron number density which have energy from $\gamma m_e c^2$ to $(\gamma+d\gamma) m_e c^2$.
The emitted power per unit solid angle coming from unit volume in the frequency range of $\omega_1\sim \omega_1+d\omega_1$ is calculated by
\begin{eqnarray}
{d P^e_{Jit}(\omega_1)\over d\Omega}d\omega_1&=& f_w
\int N_e(\gamma) d\gamma {d\hat{\alpha} d\phi\over 4\pi} {d p^e_{Jit}(\omega_1)\over d \Omega},\label{eq:defPJit}
\end{eqnarray}
where $f_w$ is volume filling factor of the regions where the waves excited by the instability occupy. 
The integration by solid angle in Eq.(\ref{eq:defPJit}) is performed to cover area B illustrated in Figure \ref{fig:radicone}. 
The integration by azimuthal angle $\phi$ is trivial since the integrand is axisymmetric against $v_z'$ axis. 
Let's define a new variable $y$ as $y\equiv 2\gamma^2\omega_0/\omega_1$. 
Since the frequency of the emitted radiation depends on $\gamma$ and $\hat{\alpha}$, 
integration variables $(\gamma,\hat{\alpha})$ are able to be transformed to $(y,\omega_1)$. 
It is performed by 
\begin{eqnarray}
d\gamma d\hat{\alpha}&=&{\partial (\gamma,\hat{\alpha})\over \partial (y,\omega_1)}dy d\omega_1
=-{1\over 4\omega_1}{d y\over \sqrt{y-1}} d\omega_1.\label{eq:ga2yo}
\end{eqnarray}  
For $\theta >1/\gamma$, $T'/T$ is rewritten as 
\begin{eqnarray}
{T'\over T}&=& {1\over \pi\sin\theta}\left({2\omega_0\over \omega_1}\right)^{1/2}{\sqrt{2-y}\over y^{1/2}}. \label{eq:TpTa2yo}
\end{eqnarray}
Since the Jitter radiation from a single electron is observed only when the emission cone of the electron points the line of sight,
the available range of $y$ is limited from 1 to 2. 
Under these considerations, the integration in Eq.(\ref{eq:defPJit}) is able to be performed and 
the emitted power per unit solid angle and per unit angular frequency emitted at $\omega_1$ is obtained as
\begin{eqnarray}
{d P^e_{Jit}(\omega_1)\over d\Omega}&=&f_w{e^2 \omega_{ce1}^2\omega_0 C_p\over 16\pi^2 k^2 c^3\sin\theta}(2+\tan^2\theta) 
\left({\omega_1\over 2\omega_0}\right)^{-{p-1\over 2}}\nonumber\\
&&\times F(p), \label{eq:dPJit}\\
F(p)&\equiv& \pi F_1\left( {1\over 2}, {p+3\over 2}, 1, -1\right)\nonumber\\
&&-{1\over 2} F_1\left({3\over 2}, {p+3\over 2}, 2, -1\right), \label{eq:defF1}
\end{eqnarray} 
where $F_1$ is the first variable of the Appell hypergeometric series and 
$F(p)$ is approximated by a linearly decreasing function from 1.2 to 0.9 as $p$ increases from 0 to 4.
By using Eqs.(\ref{eq:kmax}) and (\ref{eq:omece1}), Eq.(\ref{eq:dPJit}) is rewritten as 
\begin{eqnarray}
 {d P^e_{Jit}(\omega_1)\over d\Omega}&=&f_w b^2\left({2\over a}\right)^{p+1\over 2}\left({v_{th}\over c}\right)^{3-p\over 2} 
 {e^2 \omega_{ce 0} C_p|\cos^{p+1\over 2} \theta|\over 32\pi^2 c\sin\theta }\nonumber\\
 &&\times(2+\tan^2\theta)\left({\omega_1\over \omega_{ce 0}}\right)^{-{p-1\over 2}}
 F(p),
 \label{eq:dPJitfin}
 \end{eqnarray}
where $k_m$ for low $\beta$ plasma was used. 

To obtain the observed power, the correction of the Doppler effect due to the progressive motion of the particle toward the observer 
must be made (\cite{Rybicki1979}). 
As illustrated in Figure \ref{fig:rasen}, an interval of the arrival of successive  pulses emitted by a single electron during
one period of the helical motion 
is given by $T_A=(2\pi/\omega_{se0})(1-\beta_{0\parallel}\cos\theta)\sim (2\pi/\omega_{se0})\times \sin^2\theta$. 
Therefore, the received power is obtained by replacing $T$ to $T_A$ in Eq.(\ref{eq:pJit0}). Then, we have
\begin{eqnarray}
{d P^r_{Jit}(\omega_1)\over d\Omega}&=&{1\over \sin^2\theta}{d P^e_{Jit}(\omega_1)\over d\Omega}. \label{eq:Pe2Pr}
\end{eqnarray}
However, the average received power could be treated as  same as the emitted power
under the circumstances which we are thinking of. 
A typical frequency of synchrotron emission emitted by an electron with the Lorentz factor of $\gamma$ 
is $\nu_s\sim 0.29\times (3/4\pi) \gamma^2\omega_{ce}\sin\theta$.
Since the average strength of the ordered magnetic field in our Galaxy is a few $\mu$G (\cite{Sun2008}),
the typical cyclotron frequency in our Galaxy is $\omega_{ce0}\sim 50$Hz.
The Galactic synchrotron emission is observed in the frequency range through 100MHz to 100GHz. 
Therefore, the Galactic synchrotron emission is originated from electrons in the energy range of $\gamma\sim$ a few $\times 10^4$. 
A period of gyration motion around the ordered magnetic field for an electron with a typical Galactic Lorentz factor of $10^4$ 
is $\sim 10^3$sec.
During this period, the electron travels $\sim 10^{-5}$pc along the back ground magnetic field. 
Assume that a typical coherent length of the Galactic  ordered magnetic field is shorter than 0.1pc, 
The direction of the background magnetic field changes 
$\sim 10^{-5}/0.1=10^{-4}$radian  while the electron travels along the back ground magnetic field 
within a gyration period. 
This is almost the same value as the extension angle of the emission cone of the relativistic electron, that is $1/\gamma\sim 10^{-4}$.
It indicates that the emission cone of a single electron which includes the line of sight at some time 
drops off the line of sight after the one gyration period of $T_A$. 
In other word, the consecutive emission from a single electron with a time interval of the gyration period 
does not contribute to the observed power.  
Therefore, the Doppler shift 
effect does not significantly affect the received power and 
the averaged received power is able to be treated as same as the emitted power (\cite{Rybicki1979}). 

For a randomly oriented back ground magnetic field, the power described by Eq.(\ref{eq:dPJitfin}) is averaged over the angle $\theta$. 
The averaged power per unit solid angle per unit frequency per unit volume  is obtained as 
\begin{eqnarray}
{dj_{Jit}(\omega)\over d\Omega}&=&f_w b^2\left({2\over a}\right)^{p+1\over 2}\left({v_{th}\over c}\right)^{3-p\over 2} 
 {e^2 \omega_{ce 0} C_p\over 32\pi^2 c}\left({\omega\over \omega_{ce 0}}\right)^{-{p-1\over 2}}\nonumber\\
 &&\times F(p)G(p),
 \label{eq:avdPJit}\\
 G(p)&\equiv&\int_0^{\pi\over 2}(\cos^{p+1\over 2}\theta+\cos^{p-3\over 2}\theta)d\theta,  \nonumber
 \end{eqnarray}
 where $G(2)\sim 3.5$ and $G(3.2)\sim 2.25$.
For comparison, the averaged synchrotron power per unit solid angle per unit frequency per unit volume  under 
randomly oriented magnetic field is shown
\begin{eqnarray}
{dj_{sync}\over d\Omega}&=&{3^{{p+1\over 2}} e^2\omega_{ce 0} C_p\over 16 \pi^2 c}a(p)
\left({\omega\over \omega_{ce 0}}\right)^{-{p-1\over 2}}, \label{eq:dPsync}\\
a(p)&\equiv&{\sqrt{\pi}\over 2} {\Gamma\left({p\over 4}+{19\over 12}\right)\Gamma\left({p\over 4}-{1\over 12}\right)
\Gamma\left({p+5\over 4}\right)\over (p+1)\Gamma\left({p+7\over 4}\right)}, \nonumber
\end{eqnarray}
where $\Gamma$ is the gamma function. 
The relative power of the Jitter radiation to the synchrotron emission is therefore obtained as 
\begin{eqnarray}
{dj_{Jit}/d\Omega\over dj_{sync}/d\Omega}&\sim & f_w b^2 a^{-{p+1\over 2}}\left({v_{th}\over c}\right)^{3-p\over 2}.   \label{eq:sync2Jit}
\end{eqnarray}
It shows that the Jitter radiation power is less than the synchrotron emission power at the same frequency 
as far as  $p<3$ even if $f_w=1$.

\section{Discussion}

We have revisited the plasma kinetic instability driven by temperature gradient in a magnetized plasma
which was first examined by \cite{LE1992} and summarized the characteristics of the instability with some new results. 
We proposed that the anisotropic velocity distribution function should be decomposed into two components. 
One is proportional to the temperature gradient parallel to the back ground magnetic field.
The other is proportional to the temperature gradient perpendicular to the back ground magnetic field. 
Since the amplitude of the anisotropic velocity distribution function is proportional to
the heat conductivity and the heat conductivities perpendicular to the magnetic field is significantly reduced,  
the first component of the anisotropic velocity distribution function is predominant. 
The anisotropic velocity distribution function induced by the temperature gradient 
along the back ground magnetic field drives plasma kinetic instability
and the circular polarized magnetic plasma waves are excited. 
We showed that the instability is almost identical to the Weibel instability in the magnetized plasma 
in the weak field limit. 
However, in the case of the RL instability, whether wave vectors of modes 
are parallel to or antiparallel to the back ground magnetic field, the growth rate of
one mode is suppressed and the growth rate of the other mode
is enhanced due to back ground magnetic field.  
In the low $\beta$ plasma, one mode is stabilized and only one of the modes, $\vec{k} \parallel -\vec{B}_0$ for $n=+1$ 
and $\vec{k}\parallel \vec{B}_0$ for $n=-1$ remains unstable. 
Physical reason why the splitting of the modes due to existence of  the back ground magnetic field is caused, is unclear.

The Jitter radiation spectrum formulae when the magnetized plasma is occupied by the plasma waves driven by 
the RL instability, are deduced focusing on the low $\beta$ plasma at the first time. 
The radiation caused by relativistic electrons in the turbulent magnetic fields generated by 
the RL instability in a magnetized plasma has following distinct characteristics. 
Both synchrotron emission which is linearly polarized and Jitter radiation which is slightly circularly polarized 
but with a very small polarization degree, are observed. 
The spectrum shape of the Jitter radiation is identical to that of the synchrotron emission from the same region 
but the frequency range of the Jitter radiation is shifted toward higher frequency compared to the synchrotron. 
When the spectrum index of the relativistic electron number density is less than 3, 
the amplitude of the Jitter radiation is less than the amplitude of the synchrotron radiation 
from the same region. 
These characteristics suggest that the 
Jitter radiation originated from the magnetic turbulences generated by the RL instability 
in a magnetized plasma could be the possible physical mechanism of 
 the microwave Haze emission found by the WMAP (\cite{Fink2004}, \cite{Pl2013}). 
 The application to the microwave Haze emission is examined in the forthcoming paper.

The magnetic waves excited by the instabilities considered in this paper 
may have some link to the origin of the turbulent magnetic field 
components in the astronomical plasma.  
There are observational evidences which show the existence of the significant amount of turbulent magnetic fields.
\cite{Sun2008} showed that the Galactic magnetic field model
with the regular field plus the significant amount of the turbulent field with a Kolmogorov spectrum well fits the available radio emission data 
of our Galaxy. 
\cite{Op2012}  compiled a full sky distribution of the Faraday rotation measures of the extragalactic radio sources
and showed that  the power spectrum of the rotation measures is described by $\propto k^{-2.17}$. 
\cite{MS1996} extracted informations of the magnetic field fluctuations on spatial scale of 0.01 to
100 pc by combined analysis of the rotation measures of extragalactic sources with the power spectrum 
of the thermal electron density fluctuation (\cite{ARS1995}). 
They found that  the magnetic turbulence is well described by the Kolmogorov spectrum. 
The power spectrum of the thermal electron density fluctuations is well described by the Kolmogorov spectrum 
from 100pc down to $10^8$cm (\cite{ARS1995}).  
It has been puzzling why the interstellar plasma turbulence spectrum is well described by a single power spectrum 
from the collisional scale to the collision less scale.   
\cite{Han2004} examined the spatial energy spectrum of turbulent magnetic fields 
over the scale range 0.5-15kpc by combined analysis of the rotation and dispersion measures of 
Galactic pulsars. They found a nearly flat spectrum of  the magnetic turbulence in these scales. 
The magnetic waves excited by the RL  instabilities could be possible physical mechanism to generate 
turbulent magnetic fields in the interstellar spaces especially in collision less scale. 
However, the expected spectrum of the turbulence originated from the RL instability 
is almost monochromatic at $k\sim \omega_{ce0}/v_{th}$ 
as shown in Sec.2. 
It requires some physical mechanism of down ward and up ward cascade of the turbulent energy
if the instability is a source of the turbulence.
Although the excited waves are circular polarized magnetic waves, 
the wave lengths are too small to work as in situ accelerator for 
relativistic electrons by resonant scattering (\cite{Ohno2002}).


We would like to thank Takahiro Morishima for providing us numerical values of hypergeometric functions,
and would like to thank Nobuhiro Okabe for providing us a numerical code to deduce the dispersion relations
of the plasma kinetic instability. 
MH would like to thank Yuji Chinone for valuable discussion on the Jitter radiation. 
This work has been financially supported by the grant-in-aid for Science Research (KAKENHI numbers 21111005, 15H05743,
 26247045, 25247016)
and Brain Circulation Program (R2301)
of the Japan Science Society for promotion of Science and technology.

\clearpage

\end{document}